\documentclass{moriond}

\usepackage{wrapfig}  

\def\Journal#1#2#3#4{{#1} {\bf #2}, #3 (#4)}

\def\PRD{{\em Phys. Rev.} D}

\def\be{\begin{equation}}
\def\ee{\end{equation}}
\def\bea{\begin{eqnarray}}
\def\eea{\end{eqnarray}}

\newcommand{\Photo}{}

\begin{document}
\vspace*{4cm}
\title{TESTING GENERAL RELATIVITY FROM CURVATURE \& ENERGY CONTENTS AT COSMOLOGICAL SCALE}

\author{ ZIAD SAKR, ALAIN BLANCHARD }

\address{Universit\'{e} de Toulouse, UPS-OMP, CNRS, IRAP, F-31028 Toulouse, France}

\maketitle\abstracts{
In this work we examine what are the cosmological implications of allowing the geometrical curvature density to behave independently from the energy density contents. Using the full data extracted by Planck mission from CMB, combined with BAO and SNIa measurements, we derive, in the light of this approach, new constraints on the cosmological parameters. In particular we determine the behavior of the curvature dark energy degeneracy when allowing a varying equation of state for the latter.
We also examine whether this approach could bridge the gap recently found between the Hubble parameter value determined from CMB and that from the local universe measurements} 

\section{Introduction}

One of the prediction of the theory of inflation is that the universe is spatially flat \cite{1981PhRvD..23..347G}. The spatial curvature parameter $\Omega_k$ was found smaller than
$<10^{-4}$ when constrained \cite{Linde2008}, with very high accuracy, using the observed cosmic microwave background (CMB) fluctuations \cite{2016A&A...594A..13P} combined with other cosmological observations, like type Ia supernova (SNIa)\cite{Betoule2014} and baryon acoustic oscillation (BAO) \cite{Anderson2014}.
Observations \cite{Riess98} \cite{Perlmutter} have shown that the expansion of the universe is accelerating, leading to the existence of what is called dark energy.
This relies on the general relativity (GR) paradigm in which a geometrical tensor is set equal to the matter-energy tensor. Consequently, according to GR, for the curvature parameter of the universe $\Omega_k$ is uniquely related to its energy contents.
Therefore, testing this relation  allows to test GR at cosmological scales. This can be done by introducing on one side a geometrical curvature parameter  
denoted $\Omega_{k_{geo}}$ and the other side another parameter related to the energy contents:  $\Omega_{k_{dyn}}= 1-\sum \Omega_i$.\\
Zolnierowski \& Blanchard \cite{2015PhRvD..91h3536Z}, ZB hereafter, (see also Clarkson \cite{2012PhRvD..85d3506C}), tried, using datasets from BAO, SNIa and reduced CMB parameters, to perform this test and to analyze the relation between the two curvature parameters and dark energy. One of the main result they found was that the fiducial values $\Omega_{k_{geo}}=\Omega_{k_{dyn}}=0$ were consistent with data but a significant degeneracy exists when allowing a dark energy equation of state parameter $w$ different from the value $-1$. Yu \cite{2015PhRvD..92h3514Y} performed a similar analysis with full treatment of CMB and didn't find as much degeneracy as the former study. In the present work, we follow the same approach as Yu using newly released data, trying to set the degeneracy limits. We also examine, whether in this approach, the tension found between the two determination of the Hubble expansion parameter $H_0$ from local and distant universe probes could be alleviated.\\\\

\section{Methods }

\begin{wrapfigure}{r}{60mm}
  \centering
  \label{SN_fig}
\includegraphics[width=1.2\linewidth]{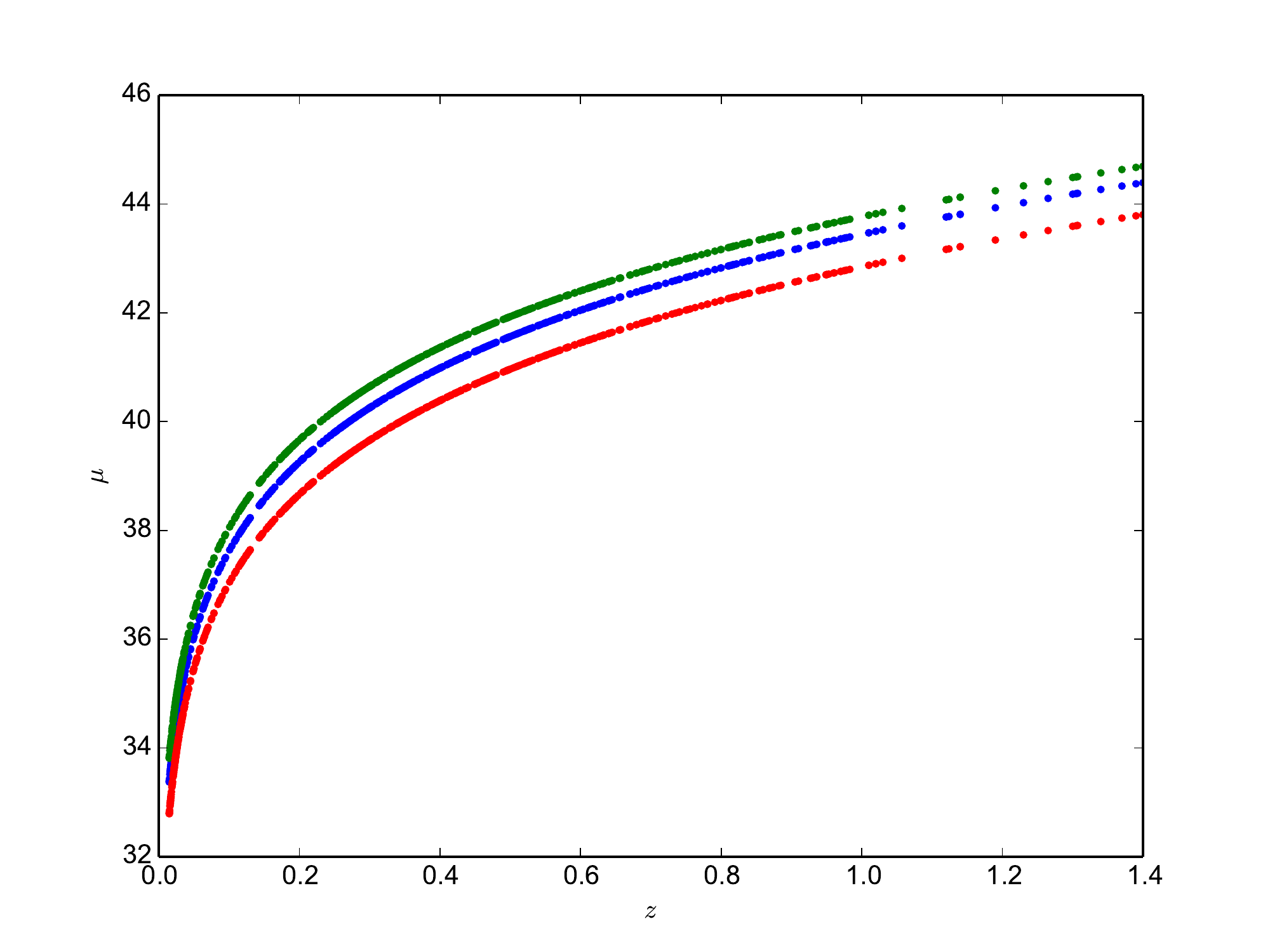}
\caption{ \footnotesize{The effect of changing the value of the parameters $\Omega_{k_{dyn}}=0.25$ (lower red line) or $\Omega_{k_{geo}}=0.25$ (upper green line)  on the SN distance modulus as a  function of redshift, compared to the fiducial value $\Omega_{k}=0$ (middle blue line)}}
 \end{wrapfigure}

To derive a solution of  GR field equations :
\begin{equation} G_{\mu\nu}= \frac{8\pi G}{c^4} \,T_{\mu\nu} \label{GR_equ}\end{equation}
for an homogenous and isotropic universe, one uses Robertson-Walker metric:
\begin{equation}
d s^2 = - c^2d t^2 +a(t)^2\left( {{d r^2}\over {1-kr^2}}+ r^2(d\theta^2+\sin^2\theta d\phi^2) 
\right) 
\end{equation}
in which $a(t)$ is the expansion factor of the universe and $k=-1,0,+1$ according  to the geometry of space.\\
That leads to the Friedman-Lema\^{\i}tre equations:
\begin{eqnarray}
\label{FR_equ}
H(z)^2 &=& H_0^2 E(z)^2 \label{eq:Hz}\\
 &=& H_0^2 \left[\Omega_m (1 + z)^3 + \Omega_{k} (1 + z)^2 + \Omega_{DE}(z)\right].\nonumber
\end{eqnarray}
where $\Omega_k=-kc^2/(Ha)^2$, which we note $\Omega_{k_{geo}}$, is the cosmological  curvature parameter  arising from the LHS of Equ.~\ref{GR_equ}. 
Equ.~\ref{FR_equ} set the equality in GR between $\Omega_{k_{dyn}}$ and  $\Omega_{k_{dyn}} = 1-\sum \Omega_i$. In the following  these two quantities are treated as independent. The evolution of the background of the universe described by Equ.~\ref{FR_equ} involving $\Omega_{k_{dyn}}$, while for deriving the angular and luminosity distance we need $\Omega_{k_{geo}}$ in addition of  $\Omega_{k_{dyn}}$. The luminosity distance for SNIa becomes:
\begin{eqnarray}
D_L  =   \frac{c \ (1+z)}{H_0 \ \sqrt{|\Omega_{k_{geo}}}|} \ {S_k} \left( \sqrt{|\Omega_{k_{geo}}|} \int_0^z \frac{ \ {d}u}{E(u)}\right). 
\end{eqnarray}
with $\Omega_{k_{dyn}}$ entering $E(z)$. The same applies for BAO probe which is determined using the angular diameter distance $D_A$. The later being related to $D_L$ through: 
$D_L(z) = D_A(z)(1+z)^2$.
We show in Fig~\ref{SN_fig} the dependence of the SN luminosity variation function of redshift on the two curvature parameters change.\\

\begin{figure}
 \begin{minipage}{0.49\linewidth}
\centerline{\includegraphics[width=0.85\linewidth]{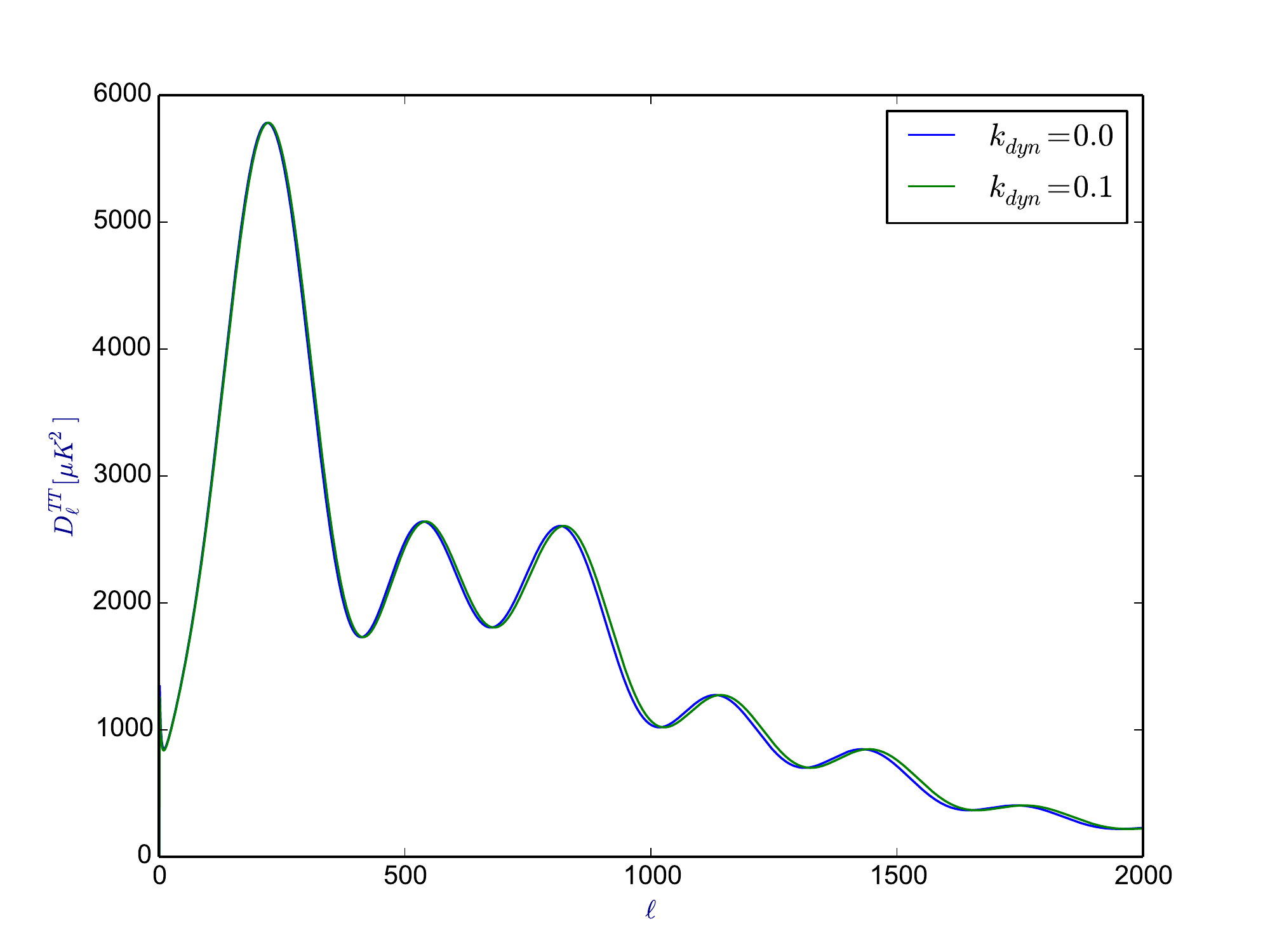}}
\end{minipage}
\hfill
\begin{minipage}{0.49\linewidth}
\centerline{\includegraphics[width=0.85\linewidth]{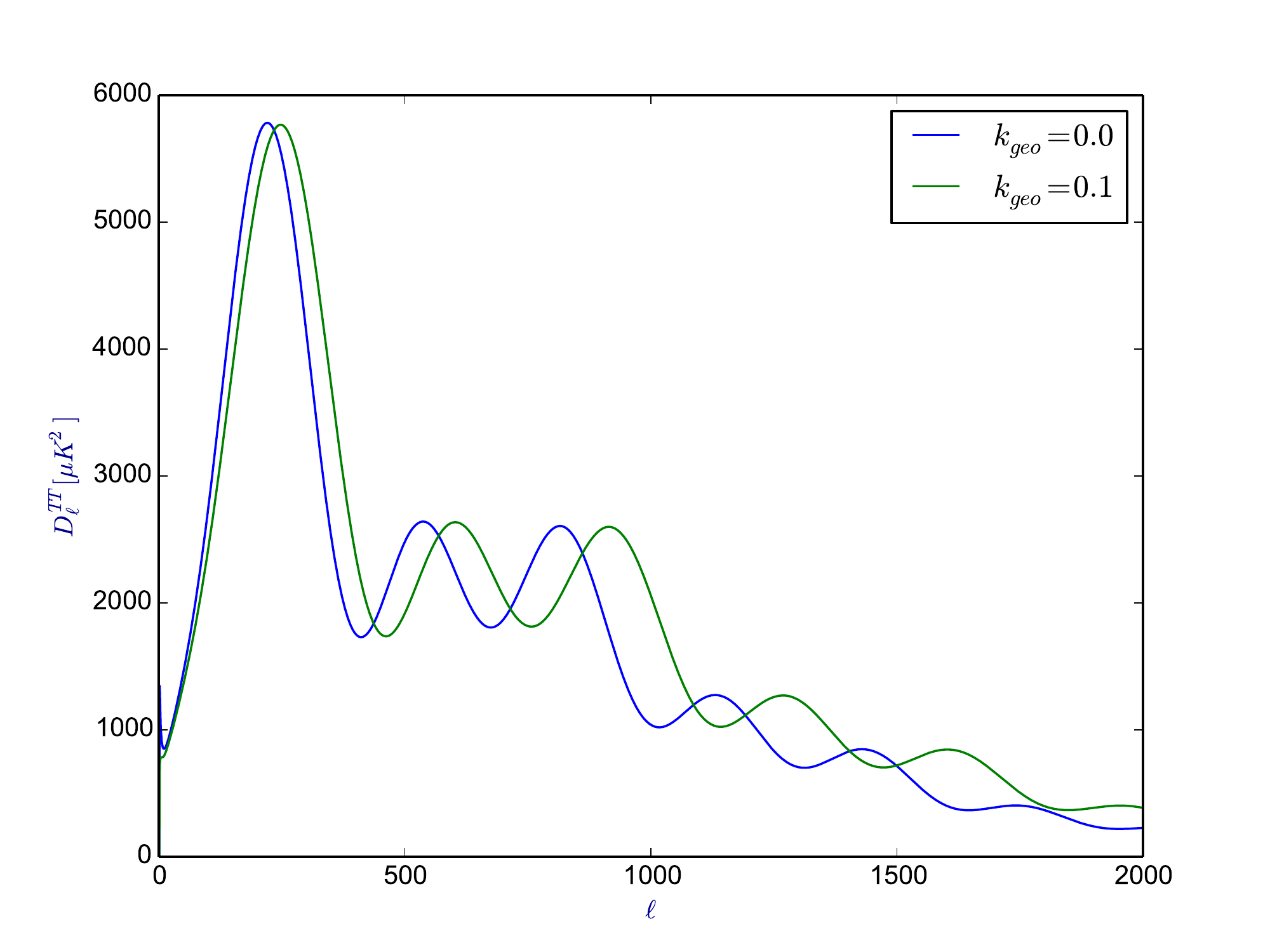}}
\end{minipage}
\hfill
\caption{The effect of $\Omega_{k_{dyn}}$ or $\Omega_{k_{geo}}$ parameters change on the angular power spectrum from CMB measurements, with the following priors $\{\omega_b,\omega_{cdm},h,A_s,n_s,\tau_{reio}\}=\{0.022,0.12,0.67,2.3e^{-9},0.09\}$}
\label{CMB_cl_fig}
\end{figure}

While for the CMB measurements, the two curvature parameters enter in the determination of the fluctuations temperature, polarization and lensing power spectrum, first with $\Omega_{k_{dyn}}$, entering the expansion and propagating in the whole derivation of the former spectrums, second with $\Omega_{k_{geo}}$ playing a significant role in different places, mainly in projections of angular correlations or distances used to derive lensing effects. 

We show in Fig.~\ref{CMB_cl_fig} the effect of changing $\Omega_{k_{dyn}}$ or $\Omega_{k_{geo}}$ parameter on the angular power spectrum from CMB measurements.

\section{Results}

To perform our analysis, we use the CMB's $C\ell{s}$ temperature, polarization and lensing measurements from Planck 2015 \cite{2016A&A...594A..13P} releases.
For the BAO probe we use datasets from Boss 2014 measurements \cite{Anderson2014} while for luminosity distance from supernovae, we use UnionII 2010 observations \cite{2010ApJ...716..712A}.

\begin{wrapfigure}{r}{80mm}
  \centering
  \label{H0_fig}
\includegraphics[width=0.5\linewidth]{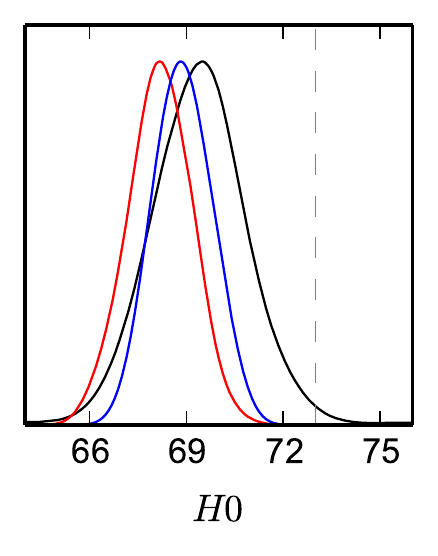}
\caption{ \footnotesize{1D Likelihood of $H_0$ (in $km$.$s^{-1}$.$Mpc^{-1}$)  using Planck, BAO and SNIa: with the standard $\Omega_k$ and $w=-1$ (blue), the same hypothesis but with $w$ free (red) and the last with  $\Omega_{k_{dyn}}$ different from $\Omega_{k_{geo}}$ with $w$ free to vary (black). The dashed line corresponds to the central value measured in the local universe by Riess et al. }}
 \end{wrapfigure}
 
To constrain our parameters we use MontePython \cite{2013JCAP...02..001A} to perform a Monte Carlo Markov Chain analysis in which the two curvature parameters assumption is implemented by changing the underlying Boltzman code and cosmological solver CLASS \cite{2011arXiv1104.2932L}. The latter can calculate the theoretical cosmological distances as well as the CMB temperature, polarization and lensing spectrums that are compared to observations in an appropriate likelihood. \\
All the cosmological parameters of the cosmological model as well as the nuisance parameters and the two curvature parameters $\Omega_{k_{dyn}}$ and $\Omega_{k_{geo}}$ are left free. The dark energy equation of state parameter $w$ is set to $-1$ in the first case and can take any constant value in the second case.

\begin{figure}
\begin{minipage}{0.33\linewidth}
\centerline{\includegraphics[width=0.8\linewidth]{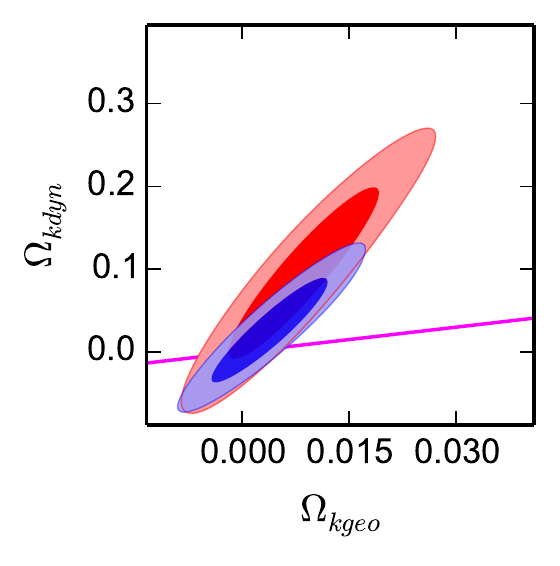}}
\end{minipage}
\hfill
\begin{minipage}{0.33\linewidth}
\centerline{\includegraphics[width=0.85\linewidth]{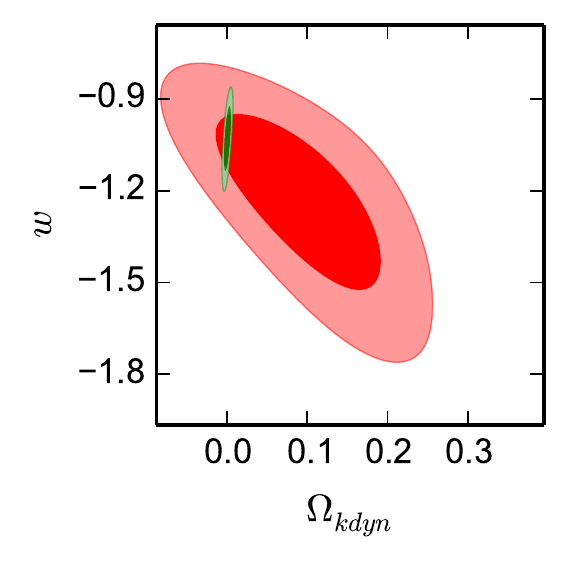}}
\end{minipage}
\hfill
\begin{minipage}{0.32\linewidth}
\centerline{\includegraphics[width=0.85\linewidth]{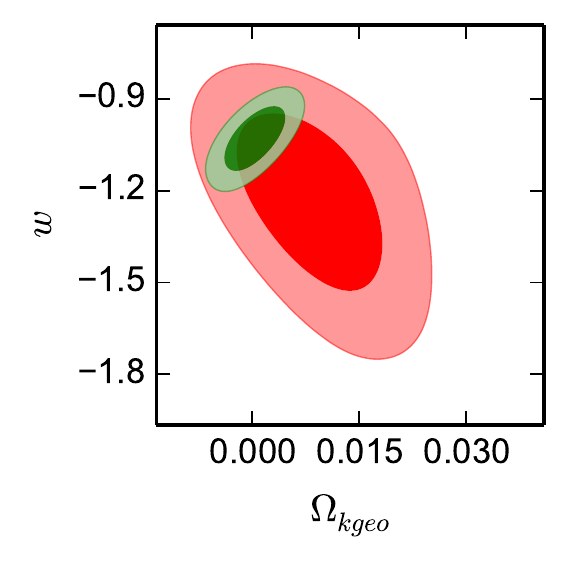}}
\end{minipage}
\caption{2D  68\%, 95\% contours for $\Omega_{k_{dyn}}$ vs $\Omega_{k_{geo}}$ (first panel), $\Omega_{k_{dyn}}$ vs $w$ (second panel) and $\Omega_{k_{geo}}$ vs $w$ (third panel), derived from MCMC analysis using CMB Planck2015 $\, TT, EE, TE$, BAO boss 2014 and SNIa UnionII. The (purple) line corresponds to GR prediction in the first panel while red CC corresponds to $w$ free to vary in all panels and blue to the fiducial case $w=-1$ in the first panel. Green CC corresponds to $w$ free vs $\Omega_k$ in GR assumption. }
\label{omk_geo_dyn_fig}
\end{figure}

Let us know examine how constraints on cosmological parameters may change when we relax the $\Omega_{k_{dyn}}$ and $\Omega_{k_{geo}}$ equality. In Fig~\ref{omk_geo_dyn_fig} we show in the first panel the confidence contours (CC) of $\Omega_{k_{dyn}}$ vs $\Omega_{k_{geo}}$ : with $w=-1$ we observe a small degeneracy but the two values are well constrained around the fiducial null value while when we allow $w$ to vary, we observe a bigger degeneracy. As can be seen from the plots of Fig~\ref{SN_fig} and \ref{CMB_cl_fig}, the variation in $\Omega_{k_{dyn}}$ is one order of magnitude higher than that of  $\Omega_{k_{geo}}$. This result agrees with both studies of ZB and Yu for that case ($w=-1$). In the second and the third panel, we consider the variation of the two $\Omega_k$ with $w$, as one can see,  $\Omega_{k_{dyn}}$ is much  degenerate with $w$, in agreement with the result obtained by ZB. One may wonder whetehr  the degeneracy of the two $\Omega_k$ primarily comes from having let $w$ free and not from having  relaxed the equality of the two curvature parameters. We examine this issue where we plot CC in both cases, i.e. in GR, i.e. with $\Omega_{k_{dyn}}=\Omega_{k_{geo}}$ (green) and without (grey). As one can the above large degeneracy results from having relaxed the equality $\Omega_{k_{dyn}}=\Omega_{k_{geo}}$.

Lastly in Fig~\ref{H0_fig}, we examine if the tension found between the determination of the Hubble parameter $H_0$ by Riess et al. \cite{2016ApJ...826...56R} from probes in the local universe and the one determined from our used probes in the more deep universe, can be alleviated. For that, we show the likelihood of the Hubble parameter $H_0$ for three cases: one with the traditional $\Omega_k$ and $w=-1$, the other with the same hypothesis but $w$ free and the last with our approach with w free to vary. We notice that the tension with $H_0$ is reduced only when we allow the two curvature parameters assumption with $w$ free while the tension remains the same within the standard picture with or without letting $w$ free to vary.

\section{Conclusions}

In this work we tested a scenario in which the curvature parameter and the energy density content parameter are distinct and vary with dark energy using the BAO, SN and the CMB measurements with a full treatment of the latter. We found that the standard picture in which $\Omega_{k_{dyn}}=\Omega_{k_{geo}}=0$ with  $w = -1$ is consistent  with present day data with $\Omega_k$ well constrained but that a significant degeneracy is observed if the equation of state parameter $w$ is allowed to vary. We also found that using this approach and letting dark energy parameter vary reduces the tension on the value of the Hubble parameter measured between local and deep universe. This shows the need of more model independent studies when analyzing cosmological observations.



\section*{References}


\begin{thebibliography}{99}

\bibitem{1981PhRvD..23..347G} Guth, A.~H., \PRD, {\bf 23}, 347 (1981)

\bibitem{Linde2008} A. Linde, \Journal{\it Lect. Notes Phys.}{738}{1}{2008}

\bibitem{Betoule2014} M. Betoule {\it et al.},
  \Journal{\it Astron. Astrophys.}{568}{A22}{2014}.
  
  \bibitem{2010ApJ...716..712A} Amanullah, R., Lidman, C., Rubin, D.,  {\it et al.}, \Journal{\it Astron. J.}{716}{712}{2010}

\bibitem{Anderson2014} L. Anderson {\it et al.}, \Journal{\it Mon. Not. Roy. Ast. Soc.}{441}{24}{2014}

  \bibitem{2016A&A...594A..13P} Planck Collaboration, Ade, P.~A.~R.,  \Journal{\it Astron. Astrophys.}{594}{A13}{2016}

\bibitem{Riess98} A.~G., Riess  {\it et al.},
 \Journal{\it Astron. J.}{116}{1009}{1998}

\bibitem{Perlmutter}S. Perlmutter  {\it et al.}, \Journal{\it  Astrophys. J.}{517}{565}{1999}

\bibitem{2015PhRvD..91h3536Z} Zolnierowski, Y., \& Blanchard, A., \PRD, {\bf 91}, 083536 (2015)

\bibitem{2012PhRvD..85d3506C} Clarkson, C., Clifton, T., Coley, A., \& Sung, R., \PRD, {\bf 85}, 043506 (2012)

\bibitem{2015PhRvD..92h3514Y} Yu, B., \PRD, {\bf 92}, 083514 (2015)

\bibitem{2011arXiv1104.2932L} Lesgourgues, J.,  arXiv:1104.2932 (2011)

\bibitem{2013JCAP...02..001A} Audren, B., Lesgourgues, J., Benabed, K., \& Prunet, S.\, \Journal{\it JCAP} {\bf 2}{001}{2013}

\bibitem{2016ApJ...826...56R} Riess, A.~G., Macri, L.~M., Hoffmann, S.~L., et al., \Journal{\it Astrophys. J.}{826}{56}{2016}

\end{thebibliography}

\end{document}